\documentclass[twocolumn]{article}
\usepackage[utf8]{inputenc} 
\usepackage[T1]{fontenc} 
\usepackage{amsmath}
\usepackage{txfonts}
\usepackage{graphics}

\date{}

\newcommand{\prl}{Physical Review Letter}
\newcommand{\pre}{Physical Review E}

\title{Clustering-Induced Attraction in Granular Mixtures of Rods and Spheres}

\author{Gustavo M. Rodríguez-Liñán, Yuri Nahmad-Molinari\\
Instituto de Física <<Manuel Sandoval Vallarta>>, Universidad Autónoma de San Luis Potosí,\\
Álvaro Obregón 64, 78000, San Luis Potosí, Mexico.\\
Gabriel Pérez-Ángel\\
Centro de Investigación y de Estudios Avanzados del IPN, Unidad Mérida,\\
AP 73 <<Cordemex>>, 97310, Mérida, Yucatán, Mexico.}

\begin{document}
  \twocolumn[
    \begin{@twocolumnfalse}
    \maketitle 
      \begin{abstract}
	Depletion-induced aggregation of rods enhanced by clustering is observed to produce a novel model of attractive pairs of rods separated by a line of spheres in a quasi-2D, vertically-shaken, granular gas of rods and spheres. We show that the stability of these peculiar granular aggregates increases as a function of shaking intensity. Velocity distributions of spheres inside and outside of a pair of rods trapping a line of spheres show a clear suppression of the momentum acquired by the trapped spheres. The condensed phase formed between the rods is caused by a clustering instability of the trapped spheres, enhanced by a vertical guidance produced by the confining rods. The liberated area corresponding to direct excluded-volume pairs and indirect depletion-aggregated pairs is measured as a function of time. The stability of rod pairs mediated by spheres reveals an attraction comparable in strength to the one purely induced by depletion forces.
      \end{abstract}
    \end{@twocolumnfalse}]
\bibliographystyle{plos2015}

\section*{Introduction}

Depletion forces, also known as excluded-volume forces, were first described in Brownian systems by Oosawa and Asakura in 1954 \cite{Oosawa1954}. They developed a theory to describe the attraction between large particles, due to depletion of small particles in the gap between the large ones. This entropic interaction governs the ordering produced in colloids by adding some depleting agent (usually a polymer or a macromolecule). Since hard particles are forbidden to interpenetrate, positions of their centers are excluded from a region surrounding the surface of a particle. Besides, since fluctuations drive the system to explore the entropy landscape, the system undergoes a segregation in size or shape. This segregation produces closely packed phases of those particles capable to liberate more volume when jointed together, creating more roaming space available to those of smaller size, which remain in a fluid phase. 

The overlap of exclusion volumes---defined as the space where the center of a particle cannot penetrate, due to the presence of another hard particle---determines the depletion zone and, thus, the strength of the depletion force. This force will lead to segregation if there is a difference among their shape, size or other physical property, such as the inter-particle coefficient of restitution or pair potential. 

Liberated volume (or area, in two dimensions) should be understood as the new region in which centers of particles are allowed to roam when exclusion regions overlap. Thus, liberated volume is defined as the excluded volume of two hard particles, considered independently, minus the excluded volume of a pair of particles in contact, taken as a single object. The entropy of those particles that liberate the largest amount of volume diminishes when they are ordered in a condensed phase; nonetheless, this condensation allows liberation of new volume, in which the other particles now can roam, increasing, thus, their entropy and the one of the whole system. Likewise, at high enough pressure or density, a hard-sphere gas undergoes a solid-liquid phase coexistence starting at packing fraction of 0.492 (0.69 for hard disks), ``where the usual glue of attractive interactions is replaced by pressure \cite{Hoover1968}.'' 

Depletion forces have been broadly investigated at colloidal scales \cite{Lekkerkerker2011}.  Intensity of depletion forces in colloids depends directly on the pressure exerted from the region not depleted of particles. This pressure, in turn, depends on the mean-square velocity of the surrounding particles or the temperature of the suspension. In this sense, increasing the temperature of the system would lead to an increment in depletion attraction. Pressure imbalance between the depleted and non-depleted regions gives rise to the effective attraction.

Depletion interactions have also been extensively studied in mixtures of viruses and colloids \cite{Adams1998} and in biological, plate-like particles---such as erythrocytes \cite{Lekkerkerker2011}---as well as in granular systems 
\cite{Bose2005,Melby2007,Sanders2004,Duran1998,Galanis2010}. Furthermore, these depletion forces can produce a layering effect at molecular or colloidal scales \cite{Israelachvili1983}, not yet observed in granular materials. 

Even though granular materials are composed of solid particles, they can flow as a liquid through an hourglass \cite{Wu1993}. They can also form planetary rings of icy grains colliding against each other \cite{Spahn1998} or even condense into piles of sand, gravel and boulders in a quiescent state---like in the case of asteroid (25143) Itokawa \cite{Fujiwara2006}. These examples represent fluid, gaseous or solid states of granular matter, respectively \cite{Jaeger1996}. Ideal granular gases are often thought as an ensemble of athermal hard spheres (Brownian motion is negligible), in which energy is injected trough the walls of the system, and dissipation of energy proceeds through inter-particle or particle-wall collisions. They ideally behave as hard-sphere gases with dissipation, and finite-volume effects of their constituent particles can be expected, as in the usual hard-sphere case. Being intrinsically dissipative systems due to inelastic collisions, granular materials are kept in a steady state by means of injection of new energy into the system. This steady state resembles an equilibrium, gaseous phase, but novel phenomena emerge from inelastic collisions. For a given granular temperature at high densities, the number of collisions per unit time among particles increases as the distance between particles decreases, and this, in turn, leads to an increment in energy dissipation rate. This phenomenon has been called clustering instability \cite{Goldhirsch1993} and produces a phase separation in agitated granular gases in experiments and simulations \cite{Olafsen1998}. These ordered clusters are kept in a condensed state due, not only to the pressure exerted by the gaseous phase, but also to the clustering instability itself, which effectively behaves as an attractive, pair-potential interaction \cite{Bordallo-Favela2009} in an equilibrium system.

In this paper we report an effective attraction between pairs of granular rods roaming in a quasi-2D sea of spheres. This attraction is enhanced by clustering, which produces an immobile chain of particles trapped between rods that behaves like a new large intruder, which, in turn, might interact with another intruder. We will refer to this phenomenon as ``indirect depletion force''. Indirect depletion forces can clearly be distinguished from standard depletion forces by the fact that the range of the latter is, at most, one diameter long, whereas indirect depletion forces keep their influence at longer distances. These indirect depletion forces, which emerge from clustering instability, are presented for the first time and could be helpful in understanding how large structures of grains form, in spite of the large mean kinetic energy of interacting particles. 

\section*{Methods}

\begin{figure*}[htb] 
\noindent\resizebox{\textwidth}{!}{\includegraphics{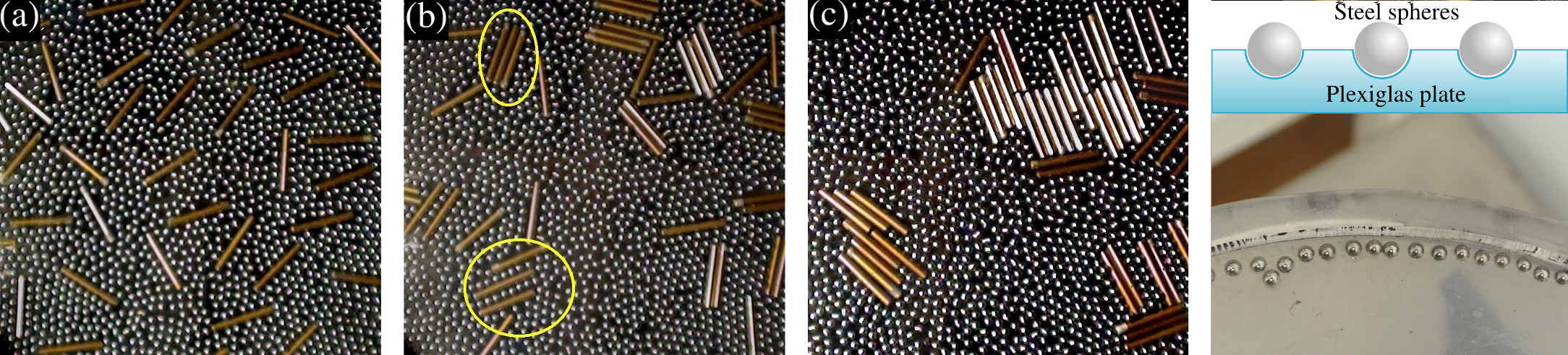}}\\
\caption{\label{System} Initial configuration of a mixture of rods and spheres (a) and its evolution after fifteen (b) and thirty minutes (c). In (d) we show a sketch of the edge of the plate with embedded spheres and a close-up to it. Highlighted in (b) are direct depletion aggregates (DDA) and indirect depletion aggregates (IDA). Rods are 23.1~mm long, and they are used as scale bar.}
\end{figure*}

Our experimental system consisted of a granular mixture composed of 50 brass rods (length $L = 23.1$~mm and diameter $\sigma = 3.3$~mm) and 2738 steel spheres, having the same diameter $\sigma$ (sphere-sphere and sphere-rod coefficients of restitution were $\epsilon_{ss} = 0.9$ and $\epsilon_{sr} = 0.8$ at a relative velocity of about 0.5~m/s), on top of a flat, horizontal Plexiglas plate (see Fig.~\ref{System}). The system was vertically shaken by a modal shaker fed with a sinusoidal signal of frequency $\nu = 60$~Hz and amplitude $A = 0.17$~mm (dimensionless acceleration, $\Gamma = A(2\pi\nu)^2/g = 2.4$, in most experiments, except where otherwise indicated). A frequency of 60~Hz was chosen because the corresponding oscillation period allows the granular gas to be excited before it relaxes forming clusters, \emph{i.e.}, the oscillation period is less than the relaxation time of the granular gas. The mixture was contained in a disc-shaped cell (diameter $D = 210$~mm and 5~mm in height) covered by a Plexiglas lid. The cell was filled at an area fraction for spheres ($\phi_s = N_s\sigma^2/D^2$, with $N_s$ the total number of spheres) of 67.6\% and an area fraction for rods ($\phi_r = N_r L \sigma/\pi(D/2)^2$, with $N_r$ the total number of rods) of 11.0\%. At the edge of the plate, close to the vertical confining wall, we embedded steel spheres, identical to those roaming freely, into the floor at a depth of $\sigma/2$, in order to transfer horizontal momentum to the rods and beads, avoiding this way fringe effects (see Fig.~\ref{System}d).

In order to characterize the slow dynamics of the system, digital video was recorded from above at 30~fps (Sony DCR-SR220). Starting from a configuration in which rods have ostensibly a homogeneous distribution on the plate (Fig.~\ref{System}a), the evolution of rods is recorded and analyzed \emph{a posteriori.} Snapshots of the system were taken every 30~s, and aggregates of rods were identified by eye. We took into account those aggregates consisting of two or more rods that met parallel or perpendicular in close contact, and pairs of parallel rods that trapped a line of spheres between them. We manually measured the excluded area of aggregated and free rods (defined as the projected area on the image where the center of a sphere cannot penetrate around a rod or an aggregate) on each image. For pairs of rods trapping spheres, we considered the line of spheres within as part of the aggregate. Afterwards, we define the liberated area of the whole system as the sum of excluded areas of all rods when they are separated (the initial configuration) minus the excluded area of solitary rods and aggregates in a given time.

In order to get velocity distribution functions and register the system's fast dynamics, we used a high-speed Red Lake Motion Meter camera running at 500~fps. By means of an ImageJ macro (consisting in adjusting brightness and contrast, thresholding and binarizing each frame), spheres were identified on each frame by finding the reflexion of a point-like light source on each sphere, obtaining this way their positions. Afterwards, consecutive lists of positions of centers of mass of particles were subtracted, carefully verifying that two rows in both lists corresponded to the same particle, and rejecting them if particles were interchanged by the macro; from this subtraction, a velocity distribution for spheres was obtained. The resolution of the images was 8.2~pixels/mm. An average of 130 spheres could be identified on each frame, whereas the total number of analyzed frames was 144. We chose those zones where a pair of rods trapped a line of spheres, and we calculated separately velocities for spheres outside the configuration and inside it. Each frame was rotated so that the pair appeared vertical. From that configuration, we could define parallel and perpendicular components of velocities, with respect to the rods, defining as positive velocities those components pointing upwards or rightwards, respectively. The orientation of the rods was automatically measured by the macro, using the reflexion of the rods, with an uncertainty of less than one degree.

In order to quantify the stability of aggregates consisting of two parallel rods trapping a line of spheres, we measured the survival time of an isolated aggregate as a function of shaking amplitude; frequency was kept constant, since the response function of the modal shaker is not linear with increasing frequency. We set this configuration as the initial condition, in this case filling the system with spheres at an area fraction of 68\% and no rods, except the pair forming the aggregate. Twenty measurements were made for each value of amplitude. An aggregate was considered broken when the orthogonal distance from one extreme of a rod to the corresponding extreme of the other is longer than three diameters (two spheres fit within) and when both rods are not parallel anymore (\emph{i.e.}, when they form an angle $> 5^\circ$). Double lines of spheres trapped among rods were not taken into account for simplicity, due to their scarcity. 

\section*{Discussion}

The mixture evolves from an initial, manually set distribution of rods having a homogeneous appearance, towards islands of aggregated rods that stick together laterally, forming dimers, trimers or polymers of rods. In Fig.~\ref{System}a, b and c, typical initial, intermediate (after 15~minutes) and final (after 30~minutes) configurations are shown, respectively.

\begin{figure}[htb]
\noindent\resizebox{\columnwidth}{!}{\includegraphics{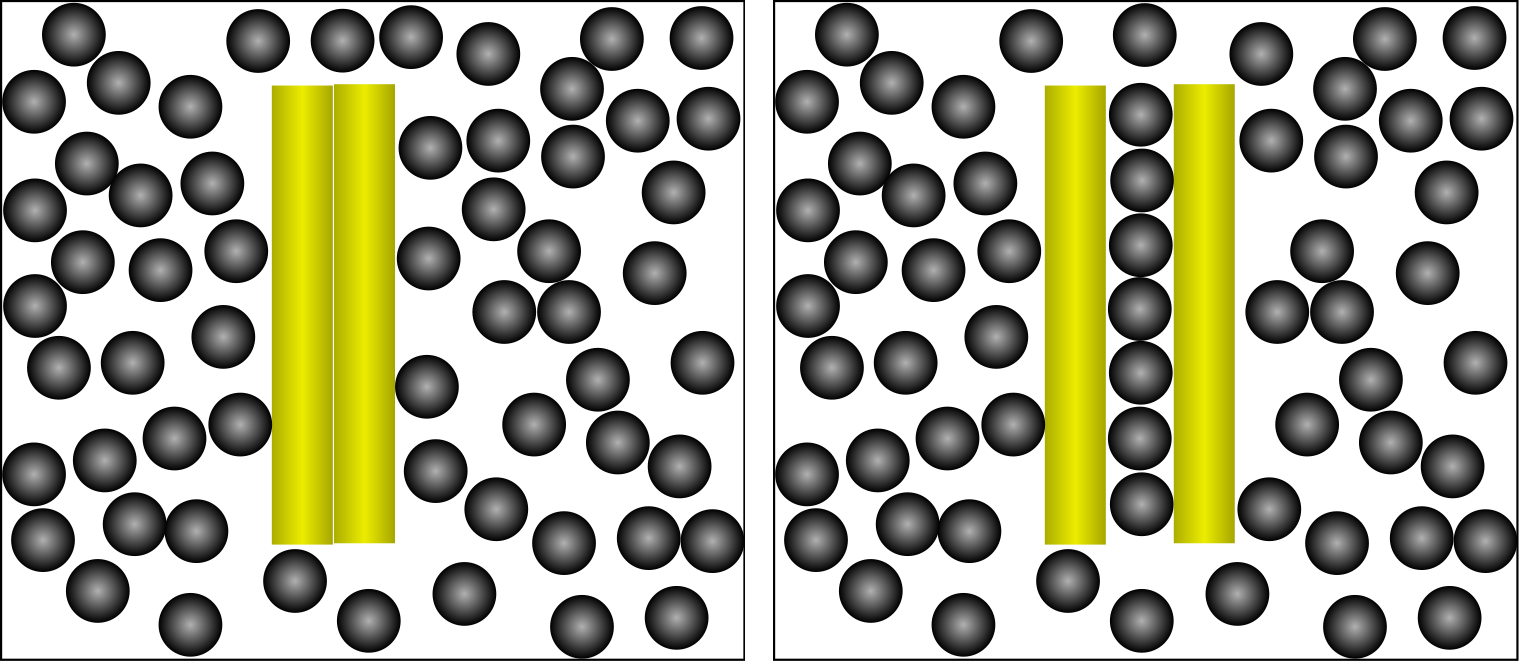}}\\
\caption{\label{IDA} (Left) Dimer formed by direct excluded-volume forces, producing a ``direct depletion aggregate'' (DDA). (Right) Configuration formed by indirect depletion forces, giving rise to an ``indirect depletion aggregate'' (IDA).}
\end{figure}

As can be seen in Fig.~\ref{System}b, a phase separation occurs, and several aggregates form after the experiment starts. The two new phases are constituted by a gas of spheres with just few rods diluted in it and several islands of aggregates of parallel rods. These aggregates can consist of two, three or more parallel rods in close contact, or two parallel rods sandwiching a chain (or even two lines) of spheres in between. Two representative examples of such configurations are highlighted on the picture. Rods that stick together in close contact will be referred as ``direct depletion aggregates (DDA)'', while configurations of sandwiched spheres will be called ``indirect depletion aggregates (IDA)'' (see Fig.~\ref{IDA}). The main difference between both kinds of aggregates resides on whether the region between two successive parallel rods is depleted or not of spheres.

The force allowing the formation of DDA dimers is depletion force, as it has been shown in previous works in 3D granular media composed by rods and spheres, as well as 2D Monte Carlo simulations \cite{Galanis2010}. This interaction arises from the fact that if two rods meet laterally, the region between them is totally depleted of spheres and, therefore, there is no momentum transfer to the rods from this ``interior'' region. Meanwhile, spheres outside are still providing pressure to the rods, pushing them against each other.

In the case of IDA aggregates, when a line of spheres finds itself trapped by chance between a pair of parallel rods, energy dissipation due to inelastic collisions and friction makes the spheres lose mobility inside this configuration. Although the region between rods is not depleted of spheres, a closer analysis reveals that their number per unit area is always higher in the interior region ($n_i = 9.5$~particles/cm$^2$) than it is outside ($n_o = 7.7$~particles/cm$^2$). This should lead, in principle, to a higher ``osmotic'' pressure from inside the IDA configuration and produce an effective force that would push apart the rods. This, in turn, should be observed as an effective repulsion of rods, whenever a set of spheres are trapped between them, making the observed configuration unstable. Instead, these structures are very likely to be produced and are stable enough to subsist in a steady state during tens of seconds once formed, when they are isolated, or even remain during the whole experiment if they get trapped between DDA structures. This high stability indicates a significant attraction between rods in such a strongly fluctuating medium.

\begin{figure}[htb]
\noindent\resizebox{\columnwidth}{!}{\includegraphics{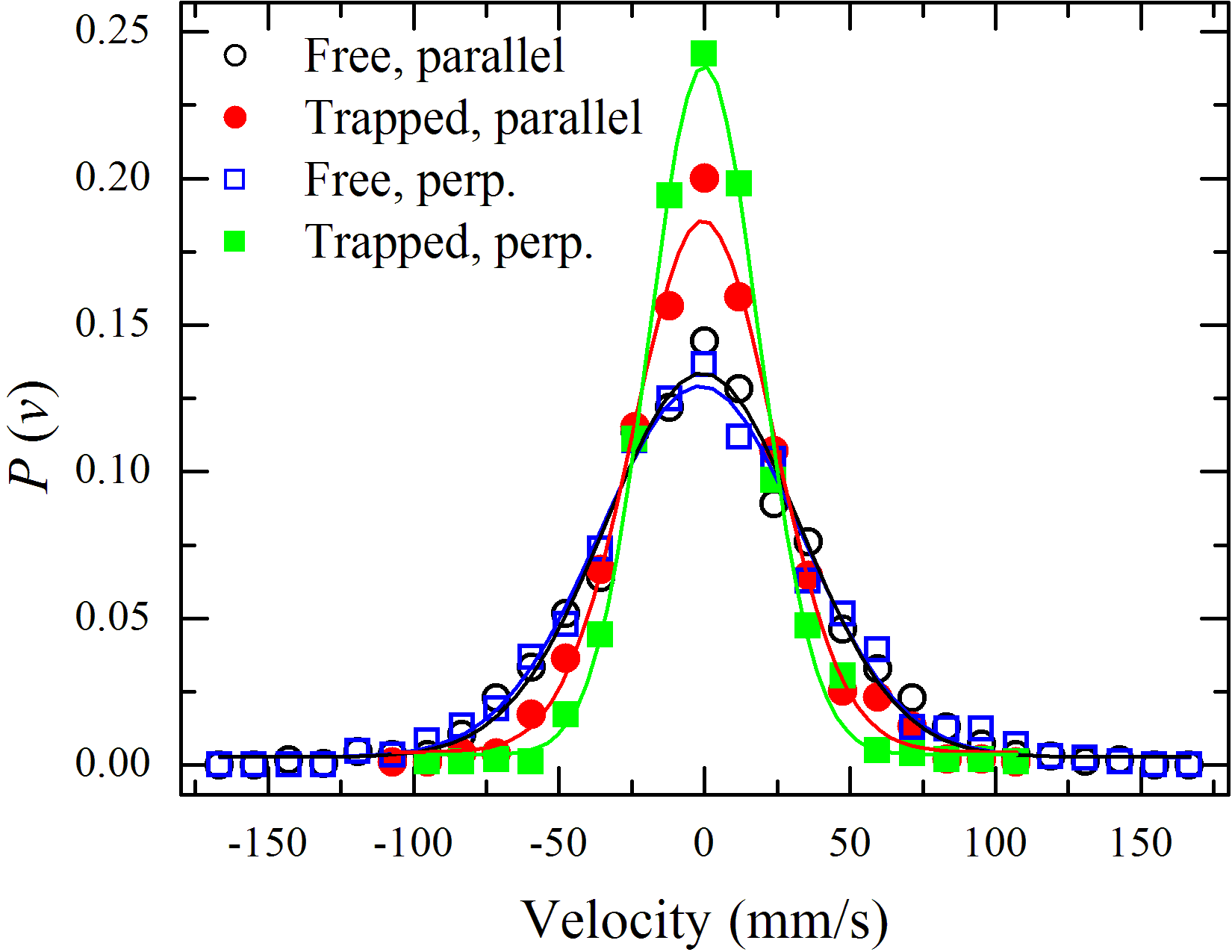}}\\
\caption{\label{Distrib}Probability distributions for velocity components along parallel (circles) and perpendicular (squares) directions to the confining rods for IDA configurations. Solid symbols represent sandwiched spheres. Hollow symbols represent free particles. Lines represent best fit to Gaussian curves.}
\end{figure}

The mechanism behind the stability of IDA configurations can be unveiled by studying the velocity distributions of the spheres inside and outside such structures. In Fig.~\ref{Distrib}, we show comparatively the resulting velocity distributions for particles sandwiched between rods and for particles in the surrounding gas. For this last case, we considered all those particles in the field of view, including those particles next to the rods and excluding only those in between. We included such spheres because, although the nearest neighbors to the rod present a lower velocity profile, they act as momentum transmitters from the spheres situated in the bulk.

A clear suppression of the perpendicular component of the motion is found for particles trapped between rods. Even more, particles sandwiched between rods are less likely to acquire momentum in the direction perpendicular to the confining rods and only receive momentum from the plate in the vertical direction and along the rods due to the constriction imposed to the spheres by the much more massive intruders (rods).

Mechanical inhibition of momentum acquisition perpendicular to the rods is due to the fact that those spheres that eventually find themselves between two rods will suffer a high number of collisions per unit time against the caging rods---rods can be thought as immobile objects given that a large fluctuation in the density of spheres should take place in order for them to have a significant displacement (their measured average velocity is $v_r = 11 \pm 4.5$~mm/s), compared with the spheres' displacement (having average speeds of the order of 50~mm/s). This phenomenon, called clustering by Goldhirsh \cite{Goldhirsch1993}, is closely related to the inelastic collapse previously discovered \emph{in silico} by  Shida and Kawai \cite{Shida1989}. This clustering cannot be measured directly in our experiments, due to spatial and temporal resolution limitations (the number of collisions per unit time diverges as rods get close each other). However, by measuring the average rod-rod distance, the average displacement perpendicular to the rods, $\bar d$ ($= 0.55\pm 0.19$~mm), can be easily inferred and, thus, the number of collisions per unit time that the spheres suffer inside an IDA. In order to do so, we fitted the perpendicular-velocity distribution function for trapped spheres (Fig.~\ref{Distrib}) with a Gaussian function and used the standard deviation of the fitting curve to obtain their mean-square velocity, $\langle v_i^2\rangle$. Dividing by the rod-rod mean distance, the collision rate is obtained ($68.7 \pm 1.8$~s$^{-1}$). On the other hand, the average collision rate among spheres outside the IDAs is obtained using the standard deviation for the perpendicular component of the velocity outside ($\langle v_o^2\rangle$) and dividing it by the mean free path $\ell$ ($=1.8$ mm). In this case, the collision rate turns out to be $38.3 \pm 1.1$~s$^{-1}$, almost one-half (56\%) of the rate inside the IDA.

The power transferred to a rod by the colliding particles inside or outside is proportional to the average kinetic energy of the particles times the number of particles that collide in a unit time. This last number can be easily evaluated by taking the product of the surface density of particles and the root-mean-square velocity in the direction perpendicular to the rods times the rod length. The number of collisions per unit length per unit time inside and outside are the densities $n_i$ or $n_o$ times the root-mean-square velocities inside or outside, \emph{i.e.}, $n_i\langle v_i^2\rangle^{1/2}$ and $n_o\langle v_o^2\rangle^{1/2}$. Consequently, the power transferred on the inner walls ($P_i$) and the outer ones ($P_o$) are given respectively by
\begin{equation}
P_i=\frac{1}{2}(1-\epsilon_{sr})m n_i\langle v_i^2\rangle^{3/2}L, \label{eqPi}
\end{equation}
\begin{equation}
P_o=\frac{1}{2}(1-\epsilon_{sr})m n_o\langle v_o^2\rangle^{3/2}L, \label{eqPo}
\end{equation}
where $\epsilon_{sr}$ is the sphere-rod coefficient of restitution, $L$ is the length of the rods and $m$ is the mass of one sphere (0.13~g for our steel spheres). In our case, the ratio of these quantities, $P_o/P_i$, is 4.9. Therefore, there will be a net attraction that grows with the mean-square velocity of particles in the gas-like phase.

The force exerted on the rod from either the inside or the outside ($F_i$ and $F_o$, respectively) is given by the power transferred, $P_i$ or $P_o$, divided by the velocity, $v_r$, the rod acquires from the colliding spheres, \emph{i.e.},
 \begin{equation}
 F_o - F_i = \frac{P_o - P_i}{v_r}. \label{eqFio}
 \end{equation}
This last velocity can be measured from a fast-dynamics analysis of the system. In the case of a dimensionless acceleration $\Gamma = 2.4$, $v_r$ turns out to be $11 \pm 4.5$~mm/s. From Eq.~\eqref{eqFio} and from the measured values of the variables in Eq.~\eqref{eqPi} and \eqref{eqPo}, we can estimate a numerical value for this force at the value of $\Gamma$. It yields $F_o - F_i = 53 \pm 23$~dyn.

Despite the fact that there is a larger number of collisions per unit time and a larger surface density of particles in the inner region (contrary to direct depletion mechanisms), fluctuations inside are so strongly suppressed that a net, positive pressure from outside develops. This confirms that the indirect depletion attraction is driven by a clustering mechanism. In other words, a chain of spheres truly condenses and behaves as a rod-like intruder.

\begin{figure}[htb]
\noindent\resizebox{\columnwidth}{!}{\includegraphics{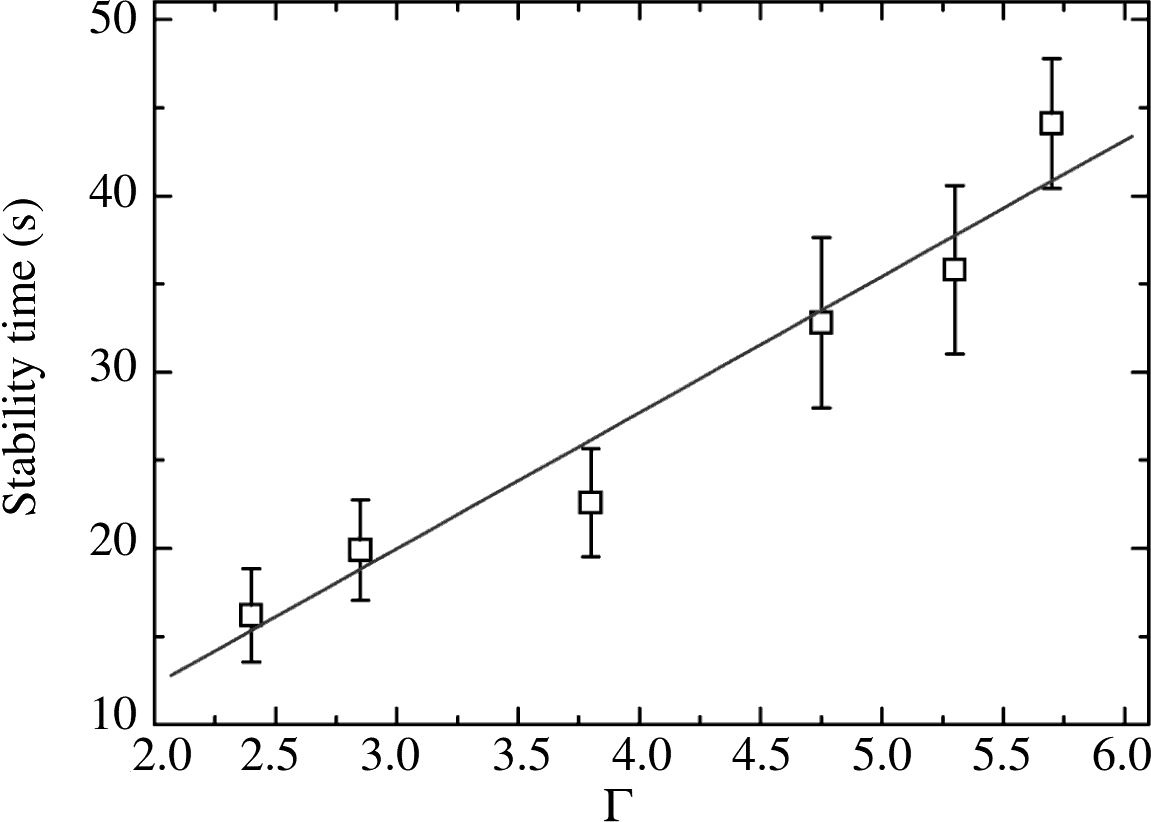}}
\caption{\label{Times} Survival times of an isolated IDA configuration as a function of dimensionless acceleration.}
\end{figure}

As stated above, for equal velocity distributions inside and outside the pair of rods, a larger number density of trapped particles within the confining rods with respect to the surrounding gas would induce an effective repulsive force between them, making the IDA configuration unlikely. Furthermore, one should expect that the stability of such configuration would decrease as the intensity of fluctuations in the surrounding gas of spheres is increased, since the rods are also subjected to these velocity fluctuations and, consequently, they would have a larger translational and rotational diffusion. On the contrary, the force between rods, and thus the stability of IDAs, increases as the shaking strength, $\Gamma_{\text{fluc}}$, does (see Fig. \ref{Times}).

The dependence of indirect depletion forces on fluctuation intensity was indirectly measured by increasing the shaking amplitude and measuring the stability (\emph{i.e.}, the survival times) of an isolated IDA configuration as the initial condition. The resulting survival times as a function of $\Gamma$ are shown in Fig.~\ref{Times}. It is worth noticing that isolated DDA dimers could last tens of minutes or even more, giving an idea of the higher stability they have in comparison with indirect depletion aggregates.

\begin{figure}[htb]
\noindent\resizebox{\columnwidth}{!}{\includegraphics{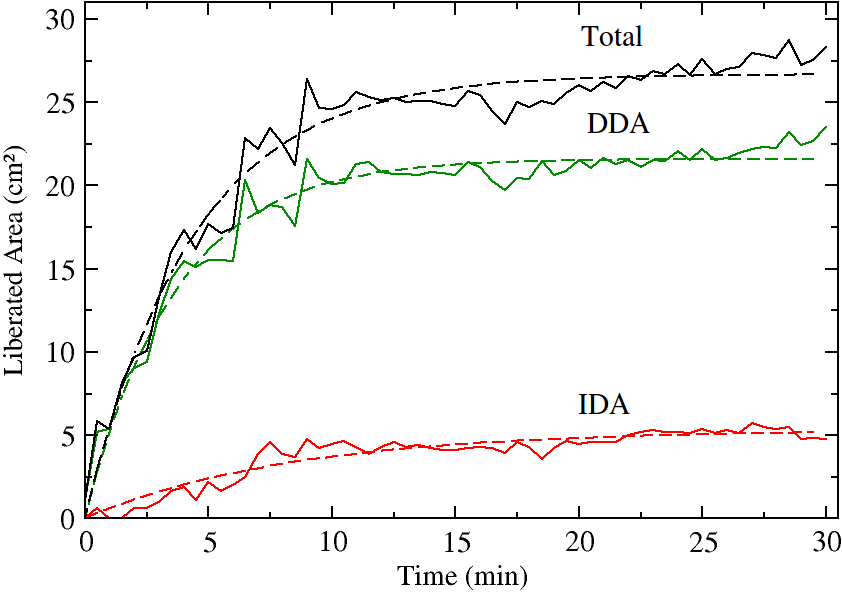}}
\caption{\label{AL}Liberated area due to aggregation of rods for DDA dimers (solid green line), IDA configurations (solid red line), and both mechanisms (solid black line) as a function of time from the start of the experiment. Dashed lines correspond to a function fitting as described in text.}
\end{figure}

For purposes of counting the effect of IDAs in increasing the area available for roaming, the overlapped area of spheres within two rods in an IDA configuration is taken into account as part of the aggregate, considering the chain of spheres as another rod. In Fig.~\ref{AL}, liberated area due to DDA structures, IDA structures, and the sum of both mechanisms is plotted as a function of time. In the same figure, dashed lines represent a fitting, corresponding to a simple model in which the number of DDA or IDA pairs formed per unit time is proportional to the number of available rods and the strength of the interaction. Under these assumptions, we fitted to these previous curves the following function of time for the liberated area $A_L$:
\begin{equation}
A_L = A_0\left[1 - \exp(-\kappa t)\right],
\end{equation}
where $A_0$ is the asymptotic liberated area and $\kappa$ is the reciprocal of a characteristic time in which aggregation phenomenon takes places, both of which are left as fitting parameters.

The characteristic times ($\kappa^{-1}$) determining the kinetics of aggregation should be proportional to the interaction strength, giving rise to faster aggregation kinetics with increasing shaking force. In our case, the ratio of these parameters for direct-depletion and indirect-depletion interaction, $\kappa_\text{DDA}/\kappa_\text{IDA}$, gave a value of 3.0. The difference among this value and the one obtained by studying the aggregation kinetics or the dynamics of particles colliding from the interior or exterior regions can be explained as follows: since IDA configurations can still provide rods to form depletion pairs, and they get stuck within depletion condensates because of dynamic arrest, aggregation kinetics in both cases are coupled. Survival times can be understood as a measurement of disassociation times for IDAs, and the characteristic time for these processes should grow linearly with increasing interaction strength, in agreement with our simple model of aggregation proposed above.

\begin{figure}[htb]
\noindent\resizebox{\columnwidth}{!}{\includegraphics{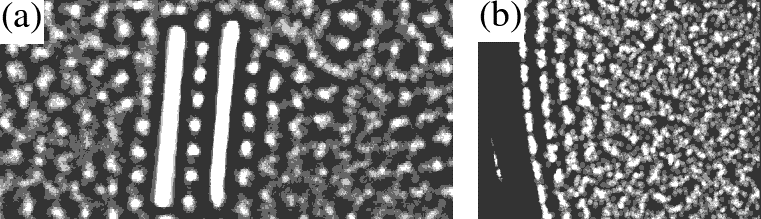}}
\caption{\label{Stack} (a) Superimposed pictures of spheres in the interior region of an IDA and its vicinity for rods of 2.31~cm. (b) Superimposed pictures of spheres in the region close to the border of the plate when the confining wall is vertical and smooth (no spheres embedded on the plate).}
\end{figure}

Boundary conditions around the surface of intruders or walls are of paramount importance to understand the formation of IDA pairs; we observe an increment in  local density around a rod or close to a wall, similar to the wetting and layering phenomenon that occurs in molecular and colloidal systems \cite{Israelachvili1983}. The wetting appearing in our experiments has also been observed \emph{in silico} \cite{Ni2015} as an increment in density close to the intruders (or walls) or as a layer of adsorbed particles kept in contact with the walls by depletion forces in a suspension of colloidal-active, hard spheres. In Fig.~\ref{Stack}, two sets of twenty superimposed pictures taken at intervals of 20~ms show this granular wetting, which appears only when the confining vertical wall and the plate are smooth; this layering does not appear when the edge of the plate has spheres embedded on it. Fig.~\ref{Stack}a depicts a typical IDA pair located at the center of the plate. In this picture, the suppression of motion perpendicular to the rods for the trapped particles can be clearly seen, as well as a gradient of mobility from the external surface of the rod towards the bulk of the gaseous phase. Such effect arises as a consequence of the zero-flux boundary condition imposed by the rods or cell walls in conjunction to the inelastic nature of rod-rod and rod-sphere collisions, leading to clustering. In Fig.~\ref{Stack}b, the superimposed pictures were taken close to the border of the plate (smooth base and vertical lateral wall) strikingly showing the layering and the increasing mobility of particle towards the center of the plate.

The density of spheres increases close to a wall or an intruder, forming a layer of low-mobility particles trapped between the gaseous phase of spheres and the surface. For colloids, a similar density profile from the wall was calculated by Fisher \cite{Fisher1964,Lekkerkerker2011}; such profile represents the onset of wetting that further develops into a true wetting or adsorbed layer if an attractive interaction exists between the wall and the particles. In our system, the latter is the case, since the inelastic character of the collisions among spheres and between spheres and rods acts as an effective attractive potential \cite{Bordallo-Favela2009}. Likewise, soft particles could be viewed as having a different effective potential and could easily present the same phenomena at a different effective radius.

The increment in density of spheres near the intruder's surface due to depletion interactions sets the onset of condensation around a rod. Clustering provokes a decrease in inter-particle distance, so the number of collisions per unit time and the dissipation rate increase. On the other hand, since particles are trapped between two massive rods, they are guided by these rods following the vertical motion transferred by the shaking system. These two phenomena will reinforce the denser phase observed between two parallel intruders.

It should be remarked that our system is far from equilibrium; thus, entropy maximization cannot be directly invoked as responsible of IDA formation. Instead, we propose Prigogine's principle of minimization of production rate of entropy \cite{Prigogine1947} as the self-organization mechanism that allows IDA formation. For instance, in epitaxial growth of granular crystals, particles nest in places that move synchronously or in phase with their close neighbors, and the nesting site correspond to the place where coordination number is a maximum \cite{Nahmad-Molinari2002}. This allows the horizontal momentum to be dissipated at the same rate as it is acquired, minimizing the rate of entropy production. Similarly, the formation process of IDAs minimizes the entropy production rate, since particles trapped between two rods synchronizes with the vertical motion of the rods. 

\section*{Conclusions} 

We have investigated a granular, two-dimensional gas composed of a mixture of inelastic, granular rods and spheres in which segregation of rods proceeds via the increment of entropy by reduction of excluded area, and the dissipation of energy leading to a clustering instability. The area available to spheres is liberated when two or more rods join side to side or when indirect depletion aggregates are created. Besides, these structures suppress the ability of the spheres trapped between the rods to acquire momentum in the direction perpendicular to the rods. This provokes a pressure imbalance between the interior and the exterior region of the configuration and leads to an effective, net attraction among rods. This is shown through velocity distributions in directions parallel and perpendicular to the rods for spheres inside and outside the IDA configuration. Finally, we show survival times of IDA configurations grow linearly with the shaking amplitude, supporting the identification of these structures as caused by a clustering instability. The ubiquity of fluctuation-induced forces in non-equilibrium systems, in particular in granular materials, could lead us to better understand aggregation and segregation mechanisms from laboratory to industrial scales.

\section*{Acknowledgments}
Thanks are given to F.\ Reyes-Tendilla for his kind help at analyzing results. This research was supported by Conacyt research grant number 221961 and Doctoral Scholarship grant number 358584.

\bibliographystyle{plos2015}


\end{document}